\documentclass[prb,letter,twocolumn]{revtex4}
\usepackage{amssymb}
\usepackage{graphics}
\usepackage{graphicx}
\usepackage{epsfig}
\usepackage{amssymb}
\usepackage{amsmath}

\begin{document}

 \title{Comment on ``Charge-parity symmetry observed through Friedel oscillations in chiral charge-density waves'' by J. Ishioka et al.}
 \author{Jasper van Wezel}
 \affiliation{Materials Science Division, Argonne National Laboratory, Argonne, IL 60439, USA}

\begin{abstract}
In their publication [Phys. Rev B, 84, 245125 (2011)], Ishioka et al. discuss the recently discovered chiral charge density wave state in {\it 1T}-TiSe$_2$ in terms of a parameter $H_{\text{CDW}}$, whose sign is suggested to correspond to the handedness of the chiral order. Here we point out that $H_{\text{CDW}}$, as defined by Ishioka et al., cannot be used to characterize chirality in that way. An alternative measure of chirality for the specific case of {\it 1T}-TiSe$_2$ is suggested.
\end{abstract}

\maketitle

It has recently been suggested that the low temperature ground state of the quasi two-dimensional, layered material {\it 1T}-TiSe$_2$ is a chiral charge ordered state.\cite{Ishioka:2010ez,vanWezel:2011wa,JvW:Physics11} In this state, the direction of the dominant component of a triple-$q$ charge density wave rotates as one progresses in the direction perpendicular to the atomic layers. In their recent publication,\cite{Ishioka:2011aa} Ishioka et al. analyze scanning tunneling microscopy images of  the chiral phase of {\it 1T}-TiSe$_2$ in terms of a parameter $H_{\text{CDW}}$, whose sign is suggested to reflect the handedness of the chiral state. Here we point out that the sign of this parameter is not well defined in {\it 1T}-TiSe$_2$, and that consequently $H_{\text{CDW}}$ cannot be used as an order parameter for the chiral state. 

The chirality parameter is defined in Ref. \onlinecite{Ishioka:2011aa} as:
\begin{align}
H_{\text{CDW}} = \vec{q}_1 \cdot \left( \vec{q}_2 \times \vec{q}_3 \right).
\end{align}
The vectors $\vec{q}_j$ in this expression represent the propagation vectors of the three components of the charge density wave in {\it 1T}-TiSe$_2$. These are well known (see for example equation (1) in Ref. \onlinecite{Ishioka:2011aa}) to be:
$\vec{q}_1 = \frac{1}{2} ( \vec{a}^* + \vec{c}^* ) $, 
$\vec{q}_2 = \frac{1}{2} ( \vec{b}^* + \vec{c}^* ) $ and 
$\vec{q}_3 = \frac{1}{2} ( -\vec{a}^*-\vec{b}^* + \vec{c}^* ) $,
with starred vectors indicating the reciprocal lattice vectors of the original lattice. From their definitions, it is immediately clear that the propagation vectors differ from their negated versions by precisely a reciprocal lattice vector.
\begin{align}
\vec{q}_1 &= -\vec{q}_1 + \vec{a}^* + \vec{c}^* \notag \\
\vec{q}_2 &= -\vec{q}_2 + \vec{b}^* + \vec{c}^* \notag \\
\vec{q}_3 &= -\vec{q}_3 - \vec{a}^* -\vec{b}^* + \vec{c}^*.
\end{align}
Since the electron momentum within the crystal lattice is defined only up to a reciprocal lattice vector, we thus find that $\vec{q}_j$ is equivalent to $-\vec{q}_j$ for all $j$. Because the sign of the parameter $H_{\text{CDW}}$ can be changed at will by adding reciprocal lattice vectors to the propagation vectors, it cannot have any physical meaning, and in particular it does not correspond to the handedness of the chiral state.

The situation can be clarified by considering the definition of the chiral charge density wave state introduced in Ref. \onlinecite{vanWezel:2011wa}. There, it is suggested that the electronic charge distribution of all phases of {\it 1T}-TiSe$_2$ can be written in the form
\begin{align}
\rho(r)&= \rho_0 +A \Re \left\{ e^{i \vec{q}_j \vec{r} } + e^{i(\vec{q}_{j+1} \vec{r} + \varphi)} + e^{i(\vec{q}_{j+2} \vec{r} - \varphi )} \right\},
\label{state}
\end{align}
where $\rho_0$ is the uniform charge distribution in the high temperature state, $j$ is either $1$, $2$ or $3$, and the other indices are defined modulo $3$. The material undergoes a transition from the high temperature uniform phase to a state with charge density modulations when $A$ becomes nonzero. This charge density wave is non-chiral as long as $\varphi=0$. Nonzero phase differences are expected to develop at a temperature $T_{\text{chiral}} < T_{\text{CDW}}$, below which the chiral sate prevails.\cite{vanWezel:2011wa} The handedness of the chiral pattern is determined by the sign of $\varphi$. This particular description of the chiral charge density wave is invoked by Ishioka et al. in their equation (1).\cite{Ishioka:2011aa}

Because both the non-chiral and the chiral charge density wave consist of a superposition of the same three propagation vectors, it is immediately clear that a parameter which only involves these vectors cannot distinguish between these states. Instead, it is essential to take into account the relative phase differences between the charge density wave components. Defining the phase relation between the components as in equation \eqref{state}, the sign of $\varphi$ directly reflects the handedness of the chiral state, and $\varphi$ may be used as an order parameter to describe the emergence of the chiral order in {\it 1T}-TiSe$_2$.

Notice however that this definition of the chiral order parameter is not universally applicable: in the related material {\it 2H}-TaS$_2$, polar rather than chiral charge order results from the charge distribution of equation \eqref{state}, and in that case $\varphi$ serves as a measure of the polarization.\cite{Guillamon:2011,vanWezel:2011qq}

Work at Argonne National Laboratory was supported by the US DOE, Office of Science, under Contract No. DE-AC02-06CH11357.

\end{document}